# Clarifying Trinko as Precedent in EHR and AI Memory Duty to Deal Cases: A New Institutional Economics Approach

By Lawrence W. Abrams | November 24, 2025

By Lawrence W. Abrams [1]

## I. INTRODUCTION

The most cited Supreme Court of the United States ( SCOTUS) antitrust opinion involving a unilateral duty to deal case is *Verizon Communications Inc. v. Law Offices of Curtis V. Trinko, LLP, (2004)* (*Trinko*).[2] Justice Antonin Scalia wrote the majority opinion. The first part is not precedent but a rhetorical warning against overreach in finding Sherman Act, Section 2 liability (§2). Scalia intended the second part to be precedent in terms of establishing the bases for a case with anticompetitive effects as not being a matter for the courts, but a matter for regulatory governance.

The *Trinko* opinion can be seen as an insightful application of New Institutional Economics (NIE) theories of polycentric governance of access rights.[3] It begins with Oliver Williamson's choice theory of the incompleteness of contracts based on relative transaction costs of alternative governance institutions. It moves on to the UCLA School of Armen Alchian and Harold Demsetz on why the nature of some goods require shared property rights for efficient exchange. This raises the question of what institutions are best to define and govern what we call the granularity of access rights.

Not mentioned in *Trinko*, but relevant as a basis for precedent, is the nature of a good in terms of the degree of excludability and rivalousness. Finally, we find support for Scalia's caution against overreach of courts in the NIE work of Elinor Ostrom and The Bloomington School on governance of common pool resources like fisheries and aquifers. They show that private "cooperative game" governance can become non-cooperative by inexplicable intrusions by legal and regulatory institutions.

By clarifying the bases for the *Trinko* opinion, we hope to reduce two distinct errors. The false positive error is citing *Trinko* as precedent when it is not. This error is so prevalent it has earned the nickname of "Trinko Creep." The false negative error is not citing Trinko when it should be. We argue below that this error will be growing in the future as *Trinko* should be precedent in cases involving regulated access rights to sensitive consumer data in electronic health records (EHRs) and Agentic AI Long Term Memory (AI-LTM).

There is another error made by commentators who unfairly compare *Trinko* to the systematic opinion of the District of Columbia Court of Appeals in the *Microsoft* case. That opinion has received much praise for clarifying the shifting burden of proof of monopolization, standards of causation, and standards for a motion to dismiss. Again, the intent of *Trinko* was not to clarify §2 law, but to clarify the boundaries of institutional governance of access rights with both pro-consumer and anticompetitive effects.

## II. The New Institutional Economics of Trinko

Scalia's Chicago School influence in the first part of *Trinko* can be found in a brazen defense of monopoly power and an application of Judge Frank Easterbrook's error-cost framework. But it is in the second part of the opinion where Scalia displays NIE reasoning to decide that the case is a matter for regulatory and not legal governance. Below is the key paragraph:

> "One factor of particular importance is the existence of a regulatory structure designed to deter and remedy anticompetitive harm. Where such a structure exists, the additional benefit to competition provided by antitrust enforcement will tend to be small, and it will be less plausible that the antitrust laws contemplate such additional scrutiny. Where, by contrast, "[t]here is nothing built into the regulatory scheme which performs the antitrust function," [caselaw citation omitted], the benefits of antitrust are worth its sometimes-considerable disadvantage.**"**[4]

NIE is especially useful in clarifying the bases for *Trinko* as precedent. It starts with the nature of the good. In the *Trinko* case, it is a physical network such that the most efficient scale would result in a single monopoly servicing a wide area. Regulation is almost synonymous with natural monopolies to ensure fair access and prevent monopoly pricing. Recognizing that there are both anticompetitive and procompetitive effects of limiting access rights, the Telecommunications Act of 1966 included an antitrust specific savings clause. Scalia mentioned its existence as evidence in support of awareness on the part of regulatory institutions of anticompetitive effects of their governance.

We argue below that *Trinko* should be precedent in cases involving personal data club goods whose access rights are defined and governed by regulatory institutions. The non-rivalousness of personal data club goods have procompetitive sharing effects. However, these goods can turn rivalrous with unauthorized access to sensitive personal data. Questions arise as to which governance institution, including private collaborative governance, would be most effective in adjudicating unilateral duty to deal cases.

Based on the nature of goods in terms of their effects on exchange and contracts, we offer the following taxonomy for clarifying the bases of *Trinko* as precedent. One key distinction is the degree of exclusivity meaning the degree by which one entity can extract value without exchange. Another key distinction is degree of non-rivalousness meaning the degree by which the consumption by one entity reduces availability to others without exchange.

**An NIE-Based Taxonomy for Identifying Trinko as Precedent**

| Excludable, Rivalous | Excludable, non-Rivalous |
|---|---|
| **Private Goods** | **Regulated Natural Monopolies** |
| Sometime non-excludable | **Regulated Club Goods** |
| involving externalities | Sometimes rivalrous due to congestion, |
| Exclusionary vertical contracts | unauthorized access, unfair distribution of gains |
|  | Unilateral refusal to deal |
|  | **Bases for Trinko precedent** |
| **Non-Excludable, Rivalous** | **Non-Excludable, Non-Rivalrous** |
| **Common Pool Resources** | **Public Goods** |
| Requiring private cooperative or public governance to prevent overuse | Requiring public governance |
| Examples - air, aquifers, fisheries | Examples - national defense, street lighting, TV and radio |

We summarize the bases by which *Trinko* should be precedent:
- Regulated natural monopolies and regulated club goods
- whose non-rivalous nature turns rivalous
- which can result in unilateral refusals to deal with both procompetitive and anticompetitive effects.

**III. Governance of Access Rights to EHRs and AI-LTM**

One economic rationale for government regulation of club goods such as electronic health records (EHRs) and Agentic AI long term memory (AI-LTM) involves a broader definition of harm to consumer welfare than higher prices and lower quantities. It is based on loss of consumer welfare from unauthorized access to personal healthcare, financial data, and other sensitive personal data. The regulation of EHR access rights by the Health Insurance Portability and Accountability Act (HIPAA) of 1966 recognized the anticompetitive effects of limiting access by including an antitrust specific savings clause.

The purpose of this section is to apply NIE to the case of *Particle Health Inc. v. Epic Systems Corporation, (Particle Health)* involving access rights to EHRs currently being litigated in the District Court Southern District Of New York.[5] We view *Particle Health* as an outstanding NIE case study how an exogenous event created a failure in access rights. This incentivized alternative governance institutions to create more granular definitions of access rights. It also would represent an interesting Bloomington School case study designed to answer the question of what factors contributed to the formation and success of a private collaborative institution for the governance of EHR access rights.

HIPPA initially defined EHR access rights very broadly as "yes" to qualified providers and "limited " to qualified payers. The passage of The Affordable Care Act in 2010 (ACA) presented a huge challenge to HIPPA definitions because it sanctioned a new payment model called value-based healthcare (VBC). VBC is heavily dependent on calibrating payments to patient outcomes. To work, VBC needs access rights to patient outcomes by so-called on-ramp companies connecting provider EHRs to validated VBC payers.

The government was slow to respond taking a full ten years for a joint public-private agreement to go live. In the interim, a group of EHR platforms and on-ramp companies formed a private collaborative institution called Carequality. Both parties to the *Particle Health* case are members of Carequality. For ten years, the private collaborative Carequality was successful in defining and governing VBC access rights. Epic Systems has approximately 60% market share of the large hospital-centered EHR platform market with estimated 2025 revenue of $6 Billion. A small part of that comes from its VBC on-ramp business. Particle Health is a startup VBC on-ramp company with projected 2025 revenue of $20 Million.

On April 5, 2024, Epic Systems cut on-ramp access to a number of Particle Health customers, notably a law firm specializing in mass tort defense and two for-profit VBC auditors hired by payers.[6]  The dispute was submitted to Carequality and settled with penalties to Particle Health on October 8, 2024. [7]  As far as we can tell no antitrust allegations were raised during these deliberations. Yet on September 23, 2024, two weeks prior to settlement, Particle Health filed a §2 case against Epic Systems claiming attempt to monopolize the on-ramp market in the District Court Southern District Of New York. To our knowledge, the court has not yet cited *Trinko*.

A week does not go by without seeing some article about a debate on regulatory governance of AI. Given that the AI-LTM market will be dominated by a few firms, access right limits will have both anticompetitive and pro-consumer welfare effects. Currently there is "nothing built into the regulatory scheme which performs the antitrust function." (*Trinko*)  On one side of the debate are the so-called "accelerationists." They advocate "incompleteness" of access rights except for broad federal regulations that preempt state laws. On the other side are the so-called "safetyists" who advocate very granular state-by-state regulations. Polycentric governance of AI would benefit by a series of early district court antitrust cases involving unilateral duty to deal. What matters most in these cases will be early clarifications of the applicable antitrust law and the corresponding bases for the opinion.

https://carequality.org/statement-regarding-conclusion-of-implementer-disputes/#:~:text=The%20Carequality%20community%20is%20a,overview%20of%20the%20process%20here